\documentclass[aip,graphicx]{revtex4-1}

\usepackage{graphicx}
\usepackage{gensymb}
\usepackage{color}
\usepackage{ulem}
\usepackage{amsmath,amssymb}
\usepackage{endnotes}
\usepackage{ragged2e}

\makeatletter

\def\enoteheading{\section*{\notesname
  \@mkboth{\MakeUppercase{\notesname}}{\MakeUppercase{\notesname}}}%
  \mbox{}\par\vskip-2.3\baselineskip\noindent\rule{.5\textwidth}{0.4pt}\par\vskip\baselineskip}
\makeatother

\def\Gammabar{$\bar{\mathrm{\Gamma}}$}
\def\Kbar{$\bar{\mathrm{K}}$}
\def\Mbar{$\bar{\mathrm{M}}$}
\def\Sigmabar{$\bar{\mathrm{\Sigma}}$}
\def\kbarp{\Kbar$^{\prime}$}

\draft 

\begin{document}


\title{Access to the full 3D Brillouin zone with time resolution, using a new tool for pump-probe ARPES} 

\author{Paulina Majchrzak}
\affiliation{Department of Physics and Astronomy, Interdisciplinary Nanoscience Center, Aarhus University, 8000 Aarhus C, Denmark}%

\author{Yu Zhang}
\affiliation{Central Laser Facility, STFC Rutherford Appleton Laboratory, Research Complex at Harwell, Harwell, OX11 0QX, United Kingdom}

\author{Andrii Kuibarov}
\affiliation{Leibniz IFW Dresden, Helmholtzstr. 20, 01069 Dresden, Germany}

\author{Richard Chapman}
\affiliation{Central Laser Facility, STFC Rutherford Appleton Laboratory, Research Complex at Harwell, Harwell, OX11 0QX, United Kingdom}

\author{Adam Wyatt}
\affiliation{Central Laser Facility, STFC Rutherford Appleton Laboratory, Research Complex at Harwell, Harwell, OX11 0QX, United Kingdom}

\author{Emma Springate}
\affiliation{Central Laser Facility, STFC Rutherford Appleton Laboratory, Research Complex at Harwell, Harwell, OX11 0QX, United Kingdom}

\author{Sergey Borisenko}
\affiliation{Leibniz IFW Dresden, Helmholtzstr. 20, 01069 Dresden, Germany}

\author{Bernd B\"uchner}
\affiliation{Leibniz IFW Dresden, Helmholtzstr. 20, 01069 Dresden, Germany}

\author{Philip Hofmann}
\affiliation{Department of Physics and Astronomy, Interdisciplinary Nanoscience Center, Aarhus University, 8000 Aarhus C, Denmark}%

\author{Charlotte E. Sanders}
\affiliation{Central Laser Facility, STFC Rutherford Appleton Laboratory, Research Complex at Harwell, Harwell, OX11 0QX, United Kingdom}%
 \email{charlotte.sanders@stfc.ac.uk.}

\date{\today}


\begin{abstract}
Here we report the first time- and angle-resolved photoemission spectroscopy (TR-ARPES) with the new Fermiologics ``FeSuMa" analyzer.  The new experimental setup has been commissioned at the Artemis laboratory of the UK Central Laser Facility.  We explain here some of the advantages of the FeSuMa for TR-ARPES and discuss how its capabilities relate to those of hemispherical analyzers and momentum microscopes.  We have integrated the FeSuMa into an optimized pump-probe beamline that permits photon-energy- (\textit{i.e.,} $k_z$-) dependent scanning, using probe energies generated from high harmonics in a gas jet.  The advantages of using the FeSuMa in this situation include the possibility of taking advantage of its ``fisheye'' mode of operation.
\end{abstract}


\maketitle 

\section{\label{introduction}Introduction}

Pump-probe time- and angle-resolved photoemission spectroscopy (TR-ARPES) presents challenges, both with respect to light sources and to detection, that do not arise in ``static'' ARPES measurements of systems at equilibrium.  In this paper, we describe the commissioning of the newly-developed ``FeSuMa'' analyser on a beamline for high-harmonic generation (HHG) at the UK Central Laser Facility's Artemis Laboratory.  We demonstrate efficient acquisition of high-quality ARPES spectra of optically pumped excitations close to the Fermi level, and we use the FeSuMa's ``fisheye'' measurement mode in combination with the beamline's capability to switch rapidly between ultraviolet photon energies that are generated as high harmonics in an Ar gas jet.  We suggest that this measurement configuration offers major benefits as a cost-efficient laboratory-scale approach to time-resolved TR-ARPES.

\subsection{\label{arpes}State of the Art in Pump-Probe TR-ARPES}

Compared to its static equivalent, an ultrafast TR-ARPES measurement adds a ``pump'' laser pulse, which---arriving at a well-defined delay time $\Delta$ before the system is probed---promotes the system into an optically excited state.  The transient, out-of-equilibrium state and its evolution in time are the subject of study.  Both the probe and the pump pulses must be short (\textit{i.e.,} must have narrow width in the time domain) relative to the time scales of the physical phenomena to be measured, and the pulse train of the pump must be well synchronized to that of the probe.  The pulse widths and the pulse synchronization determine the limits of time resolution in the experiment.

\subsubsection{\label{lightsources}Background: Light Sources}

Pump-probe methods require the simultaneous generation of synchronized pulse trains of very different energies.  In TR-ARPES, an infrared (IR) or visible pulsed beam is needed for excitation, while an ultraviolet (UV) or extreme ultraviolet (XUV) beam probes the system \textit{via} photoexcitation.  The Fourier limit  places a strict boundary on the energy resolution achievable in an experiment, depending on what time resolution is needed (or, vice versa, if energy resolution is critical, then the Fourier limit determines the maximum achievable time resolution). The demands on the resolution are set by the time and energy scales of the physical phenomena of interest: for example, studies of electron-electron interactions typically demand time resolution of no worse than tens of fs \cite{Petek1997}.

Light sources commonly used to supply the pulsed ultraviolet probe beam include tabletop laser setups \cite{Zhou2018,Suzuki2021} and free-electron lasers \cite{Zhou2018,Grychtol2022,Kutnyakhov2020}.  This paper deals with tabletop laser setups.  Wavelengths down to approximately 115~nm (\textit{i.e.,} energies up to approximately 11~eV) are achievable with commercial off-the-shelf laser systems \cite{Peng2018}.  However, even at 11~eV, one can access only a relatively small section of momentum space up to approximately 1.3~Å$^{-1}$.  Off-the-shelf laser systems cannot generate photons in the tens-of-eV range that allows ARPES to access the full three-dimensional (3D) Brillioun zone of most crystalline materials, or that gives access to shallow-lying core-level states.  To reach this range in a tabletop setup, one typically relies on HHG---usually in a gas jet\cite{Suzuki2021,Frassetto2011}.

It is possible to use a single powerful laser to generate both the IR pump and the HHG XUV beam. An advantage of this approach is that the pulse trains of the two beams are automatically synchronized.  The method works by taking the IR beam from a commercial laser system and splitting it into two parts, one of which is used to generate HHG, and the other of which is sent along a separate beam path for use as a pump.  A movable delay stage in one of the beamlines (typically the pump beamline) controls the pulse separation $\Delta$, and then the two beams are recombined.

In a 3D-dispersing system, the probe photon energy determines which part of the 3D Brillouin zone is measured.  Control of the probe energy also allows for optimization of photoemission intensity in electronic states of interest, via control of final state and matrix element effects \cite{Damascelli2003,Sobota2021, Boschini2020, Gierz2011,Gierz2012}. While this is the basis for photon-energy-dependent synchrotron-based studies of out-of-plane-dispersing ``$k_z$'' states, the situation is more challenging for laser-based experiments:  while a synchrotron (or FEL) undulator can generate probe photons with continuously tunable energy across a wide range \cite{Peatman1997}, no such continuous spectrum is possible with laser-based HHG.  Rather, HHG produces a ``frequency comb'' of odd-ordered harmonics \cite{Jaegle2000} (Fig. \ref{fig1}(b)).  A single frequency from the comb can be selected with a combination of reflective and transmissive optics \cite{Puppin2019}.  At Artemis, we take an alternative approach, using a grating monochromator, which spatially separates the frequencies of the HHG comb into a ``fan'', and a slit that picks out a single frequency from the fan \cite{Frassetto2011,Heber2022}.  When the monochromator is properly aligned, any frequency in the comb can be quickly selected on-the-fly, which provides great flexibility to choose different photon energies\cite{Frassetto2011}.  The power of this type of approach has recently been demonstrated \cite{Heber2022}.

\subsubsection{\label{analyser}Background:  Photoelectron Detection and Analysers}

The best-established technology for photoelectron spectroscopy is the hemispherical analyser (HA). This tool measures photoemission intensity as a function of momentum and energy across a wide range of binding energies and with energy resolution that is better than 1~meV \cite{Iwasawa2017}.  The HA has been the workhorse of the photoemission community, and is likely to remain so for the foreseeable future.  Moreover, state-of-the art HA technology increasingly incorporates advanced features; for example, spin detection.  However, if electronic states of interest do not correspond to a single set of emission angles along the slit direction of the analyzer, then multiple HA measurements must be taken---either by rotating the sample in front of the analyzer, or by applying ``deflector'' voltages to the electron lens column in order to sample emission angles away from the slit direction.  This works well, but can be a challenge in the context of pump-probe measurements, where each data set is intrinsically time consuming (on account of the need to acquire spectra at multiple time delays).  When important physics arise simultaneously in multiple parts of the Brillouin zone, or when photon-energy-dependent measurements will cause the 3D Brillouin zone to shrink and expand on the detector, an alternative approach to detection is desirable.  Moreover, in the case of short-pulse pump-probe applications, the high energy resolution of the analyser greatly exceeds the Fourier limit of the short light pulses.

Recent years have seen rapid advancements in new types of analyser technologies \cite{Schonhense2015,Maklar2020,Tusche2020}.  Some are based on time-of-flight (ToF), among which are various types of photoemission electron microscopes (PEEMs) and momentum microscopes (MMs) \cite{Suga2021}.  These new techniques are powerful, permitting sophisticated momentum-space and real-space mapping, but they also present new challenges.  In the case of PEEM and ToF-MM, a large potential difference must be applied between the sample and the objective lens of the electron optics.  Because these are close together (on the order of several mm), there is a possibility of dielectric breakdown across the small vacuum gap and, as a result, sample damage.  Another challenge, in the case of pulsed-probe measurements, arises with space charge effects in the electron optics \cite{Kutnyakhov2020}.  This latter issue is discussed below.  

Very recently, a new type of simple, economical, and yet highly effective photoelectron analyser has been developed.  The working principle---similar to that of velocity map imaging \cite{Eppink1997,Stei2013}---has been described in a recent paper \cite{Borisenko2022}.  From the point of view of pump-probe measurements, we find that this new analyser, which is available commercially under the name``FeSuMa'' (``Fermi Surface Mapper''), offers advantages over other types of analysers for some of the most commonly required types of pump-probe ARPES measurements.  The technology offers an efficient approach to measurement of the full Brillouin zone.  It has a very straightforward application to states near the Fermi surface, which are the primary states of interest for many pump-probe ARPES studies; however, it can also probe deeper-lying states, including shallow-lying core levels.  When used on a beamline with photon-energy control, it offers an efficient method for 3D measurements of $k_z$-dispersing states.  Additionally, its user-friendly operation and compact profile make it easy to incorporate into crowded lab spaces and into vacuum chambers that might contain several other tools for various other types of measurements.  In the sections that follow, we describe how these advantages have been integrated into our tabletop HHG beamline at Artemis to take advantage of the special capabilities of the FeSuMa in the context of pump-probe ARPES measurements.

\section{\label{setup}Experimental Details}

\subsection{\label{optical}Optical Setup}

\begin{figure}
\includegraphics[width=1.0\textwidth]{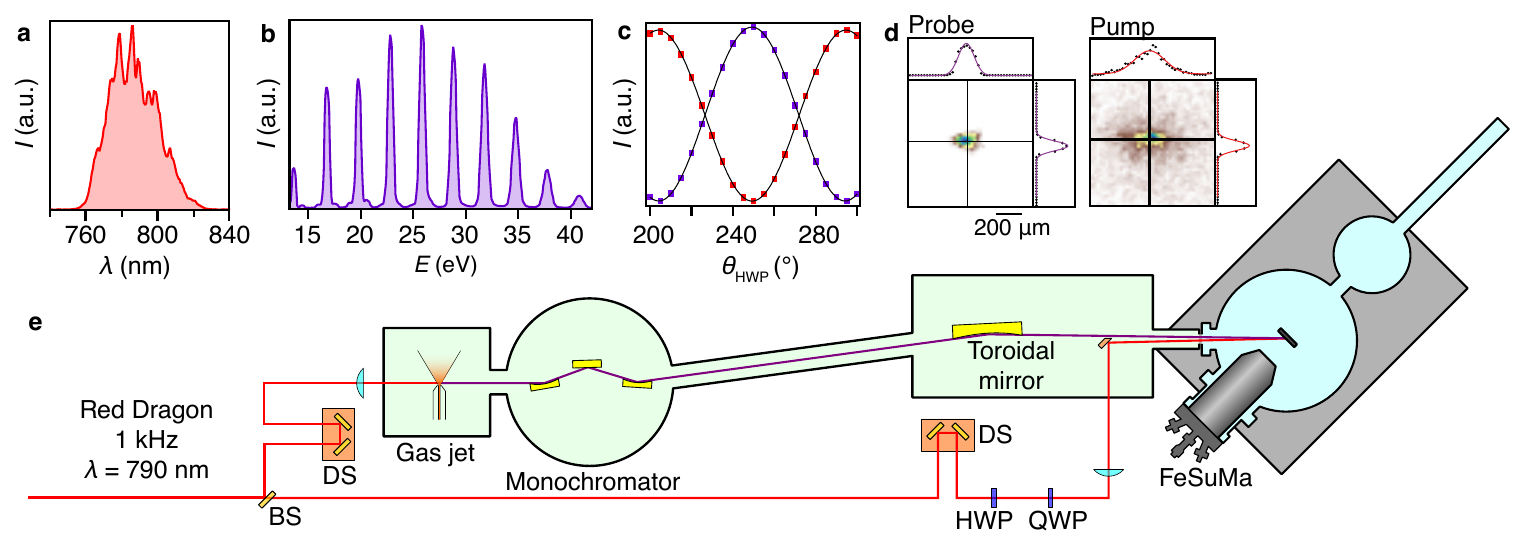}\\
\caption{(a) The spectrum of the Ti:Sapphire laser, centered at about 790~nm with a bandwidth of approximately 50~nm. (b) High-harmonics spectrum generated in the argon gas jet.  The maximum photon flux is approximately $10^{10}$~photons/second/harmonic at
27~eV. A time-preserving monochromator is used to choose between the probe energies. (c) Calibration curves for the incident pump polarisation, acquired as measured intensity through a polarizer after the QWP as the HWP is rotated. (d) The spot sizes of the probe (80~$\mu$m, FWHM) and pump beams (250~$\mu$m, FWHM), captured by FeSuMa. (e) A simplified schematic of the experimental setup.  The locations of delay stages are labeled ``DS,'' while ``BS'' indicates an 80-20 beam splitter.}
\label{fig1}
\end{figure}

In Fig. \ref{fig1}, we present the layout of the Artemis optical setup for the tests described here.
The pump and probe pulses are generated from the output of a 1-kHz Ti:Sapphire laser (upgraded from the RedDragon, KMLabs) with a pulse energy of about 3~mJ at 790~nm. The bandwidth is approximately 50~nm, as shown in Fig. \ref{fig1}(a). The output laser beam is split into two parts, with 80{\%} focused onto a 200~$\mu$m Ar gas jet, via a lens of 500-mm focal length, for HHG. An example of the resulting XUV spectrum---\textit{i.e.} the frequency comb---is shown in panel (b). For these conditions, the usable high-harmonic energies range from approximately 17~eV to 45~eV, with a maximum photon flux of about 10$^{10}$ photons/second/harmonic at 27~eV. A single harmonic is selected by a time-preserving grating monochromator\cite{Frassetto2008}. To avoid space-charge effects, which are discussed in Section \ref{spacecharge}, the photon flux was reduced to 10$^{8}$ photons/second/harmonic by an adjustable slit after the monochromator.

The remaining 20 percent of the output beam is used for pumping, either at its fundamental wavelength or after frequency-doubling or -quadrupling by beta barium borate crystals. A delay stage in the pump beamline enables time-resolved measurements. A half-wave plate (HWP) and a quarter-wave plate (QWP) are added into the pump beamline for polarization control: Fig.~\ref{fig1}(c) shows calibration data for the HWP rotation angle (QWP angle was held fixed). The pump beam is finally focused on the sample, using a lens with focal length of 1.5~m. The pump and probe beams reach the sample almost collinearly, with an angle of 45{\degree} relative to the sample normal when the sample is at normal emission relative to the detector. The pump fluence is about 2~mJ/cm$^2$, with pump spot size of approximately 250~$\mu$m at FWHM (450~$\mu$m at 1/e$^2$ width). The XUV spot size of about 80~$\mu$m is measured roughly by the size of the spot on a scintillating crystal, and confirmed by the FeSuMa in Direct Mode (see Section \ref{fesuma}). The images of the two beam spots recorded by FeSuMa are presented in Fig.~\ref{fig1}(d).
The time resolution is determined from the auto-correlation spectrum, as shown in the Supplementary Material \cite{suppMat}.

\subsection{\label{fesuma}Working Principles of FeSuMa}  The FeSuMa is a new type of ARPES analyser that combines Fourier electron optics with retarding field techniques \cite{Borisenko2022}. The lens of the device consists of several cylindrical elements that represent the simplest element of electron optics---the Einzel lens. It focuses parallel electron beams, originating from the sample surface, into corresponding points in the focal plane. This is similar to the action of a convex optical lens which makes a Fourier transformation of light. The novelty of the approach is in placing the detector, a multichannel plate (MCP), directly in the focal plane, and applying a retarding potential, $V_r$, to the front of the MCP. In practice, the focal points lie not on a plane but on a curved surface, and the detector is placed so as to achieve a reasonable balance between angular acceptance and angular resolution. The signal is amplified by a pair of MCPs in ``chevron" geometry, and is converted into photons by a phosphorus screen. A camera outside the vacuum captures the image and sends it to the computer for further processing.

\begin{figure}
\includegraphics[width=1.0\textwidth]{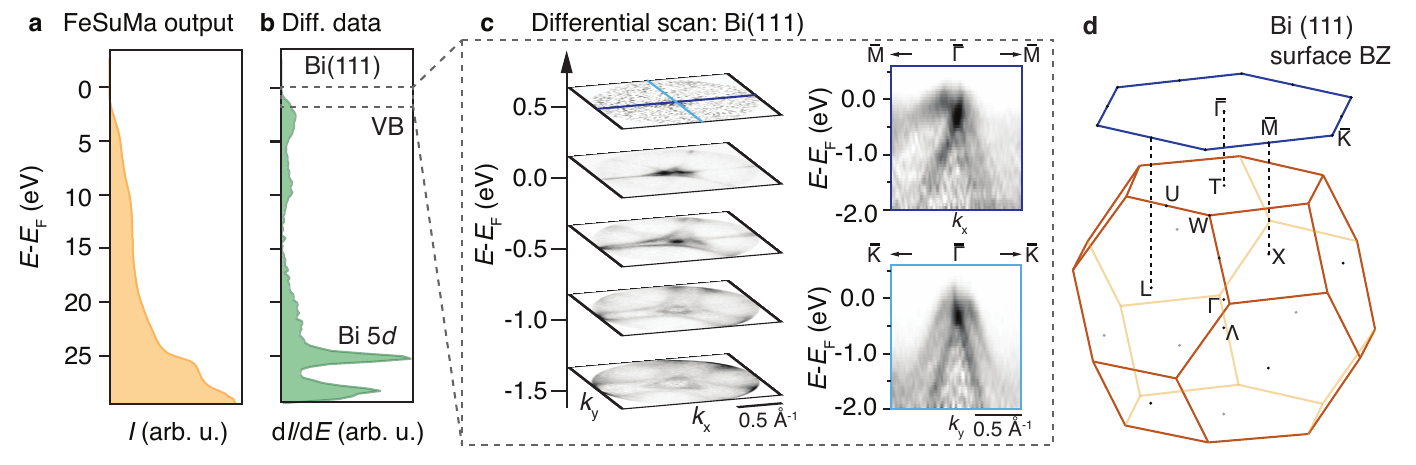}\\
\caption{(a) Example of static angle-integrated raw data obtained from Bi(111) by scanning the retarding potential $V_r$ ($h\nu$ = 37.4~eV, sample measurement temperature 300K). (b) Same data as in (b), after differentiation.  The core-level spectrum of Bi(111) is now well resolved. (c) Example of an ($E$, $k_x$, $k_y$)-resolved data set, acquired without optical pumping from Bi(111) ($h\nu$ = 22.4~eV, sample measurement temperature 78K).  The directions of the cuts correspond to the directions of the lines (matched in colour to the frames of the two spectra) in the constant-energy contour at left.  Note the scale bar at the bottom:  high-symmetry points {\Kbar} and {\Mbar} are outside the range shown in the panel.  The asymmetry of intensity in the cut along {\Mbar}-{\Gammabar}-{\Mbar} arises from matrix element effects that can be seen clearly in the constant energy slices at left. (d) Schematic of the Bi BZ, with high-symmetry points labeled.  Panel (d) is adapted from Ref.~\cite{Hirahara2006}.}
\label{fig2}
\end{figure}

By setting $V_r$ such that only Fermi-level electrons can reach the detector from an unpumped sample, one can observe the Fermi surface map directly on the screen. In order to obtain information about electrons with higher binding energies, $V_r$ is reduced step-by-step while the detector collects the integrated signal. Subsequent differentiation results in a conventional photoemission spectrum. An example of such a measurement is in Fig.~\ref{fig2}(a), where we show a Bi core-level spectrum acquired from the Bi(111) surface\cite{suppMat}. The spin-orbit splitting in the Bi~$5d$ doublet is well resolved when the spectrum is differentiated, in Fig.~\ref{fig2}(b). In like manner, to obtain the intensity distribution of a photoemission signal from valence states as a function of momentum and energy, a three-dimensional data set is recorded and then differentiated along the energy axis across a smaller range of energies close to the Fermi level. We show the example for the case of Bi(111) in Fig.~\ref{fig2}(c), where the Fermi surface, momentum distribution, and underlying dispersion of the electronic states are visible.  Due to the semimetallic nature of bulk Bi, the photoemission intensity at the Fermi level is dominated by surface states \cite{Koroteev2004}.  The bulk and surface Brillouin zone (BZ) of Bi(111) is provided in Fig.~\ref{fig2}(d) for reference.

The FeSuMa operates in three regimes: Fourier Mode, Direct Mode and Optical Mode. Within the first of these regimes, there are actually three settings, characterized by angular acceptances of $\pm 8 ^\circ$, $\pm 14 ^\circ$ and $\pm 16 ^\circ$. Angular acceptance in the Fourier modes can be extended by applying a bias potential (see discussion below and Fig. \ref{fig5})---a technique that is also used in conventional ARPES \cite{Gauthier2021}. The FeSuMa's ability to instantly detect the angular distribution of intensity allows the parameters to be quickly adjusted, minimising the distortion of the electric field caused by any non-cylindrical symmetry in the sample environment.

In the Direct Mode, the lens projects an image of the electron source in real coordinates; thus, it can be used to characterize and track the beam spot in two dimensions (see Fig.~\ref{fig1}(d)). This is a significant advantage in comparison with conventional HAs, where only one spatial coordinate, corresponding to the direction along the entrance slit, is accessible.
Since the MCP is sensitive to UV photons, Direct Mode can also be used to detect reflected or scattered light from surface features and sample edges, and thus either to track the position of the photon beam or to find flat portions of the surface (since no photons should enter the analyser from a flat sample region if the electron signal is optimised).

We finally mention here an advantage of the FeSuMa for pump-probe experiments: unlike in HAs and MMs, electron trajectories in the FeSuMa (being an order of magnitude shorter) do not pass through auxiliary focal planes or crossing points. In HAs, there are two imaging planes and one crossing point where electron trajectories are brought together (\textit{e.g.,} Ref.~\cite{Zouros2002}), and Coulombic electron-electron interactions are presumably enhanced at such points.  It is generally desirable to avoid such space charge effects, as they degrade angular and energy resolution.  In the case of MMs, electron-electron interactions both inside the focusing column and in front of the objective lens are complex and problematic \cite{Maklar2020,Schonhense2018,Kutnyakhov2020}.  The FeSuMa's design, which reduces the effects of space charge inside the electron optics, is beneficial to pump-probe measurements.  This will be discussed further below.

\section{\label{pop}Proof of Principle Data}

\begin{figure}
\includegraphics[width=1.0\textwidth]{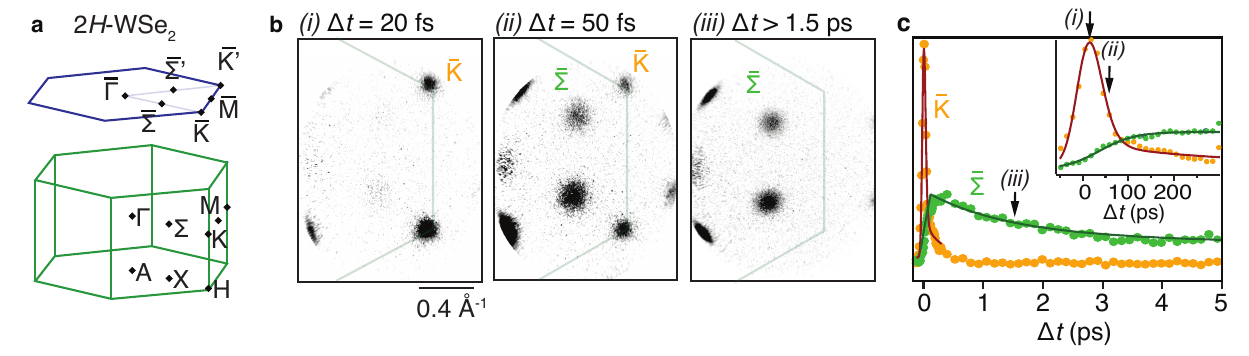}\\
\caption{(a) Surface (dark blue) and three-dimensional (green) Brillouin zones of 2$H$-WSe$_2$.
(b) ($k_x$, $k_y$) slices corresponding to different stages of the ultrafast evolution of the system after pumping with $s$-polarized light at 800~nm (probe photon energy $h\nu=22.6$~eV, sample temperature 78K).  The sample is rotated such that {\Gammabar} is at the left edge of the detector.  The {\Kbar} and {\Sigmabar} points of the Brillouin zone are labeled. (i) Just after the optical excitation, only \Kbar-points are populated. (ii) Within 50~fs of the optical excitation, electrons can be seen to transfer from the \Kbar-valleys to \Sigmabar-valleys.
Retarding potential is set such that $E-E_F = 0.65$~eV.
(iii) At longer times after the pump arrival, all of the excited electronic population has either transferred to the \Sigmabar-points or has already relaxed fully.
(c) Ultrafast dynamics of 2$H$-WSe$_2$:  orange (green) markers denote photoemission intensity difference, relative to pre-pumped intensity, integrated over the \Kbar (\Sigmabar)-points of the Brillouin zone.  These temporal dynamics are in good agreement with previously published results \cite{Bertoni2016, Maklar2020}.}
\label{fig3}
\end{figure}

In the following, we summarise the versatile applications of the FeSuMa analyser when coupled with a pump-probe setup. To facilitate comparison with similar approacheas involving HAs and MMs \cite{Maklar2020}, we benchmark the capabilities of the system using a widely studied layered transition metal dichalcogenide, cleaved bulk trigonal prismatic tungsten diselenide (2$H$-WSe$_2$). Bulk WSe$_2$ is an indirect bandgap semiconductor \cite{Hsu2017} with a hexagonal BZ that is sketched in Fig.~\ref{fig3}(a). Its valence band maximum (VBM) is located at the $\Gamma$-point, and the conduction band minimum (CBM) at the $\Sigma$-valley, in between $\Gamma$ and $K$. Upon optical excitation with a circularly polarized infrared pulse, the material exhibits spin-, valley-, and layer-polarisation \cite{Bertoni2016}.

We start by demonstrating a simple approach to a common (but historically challenging) application of TR-ARPES:  namely, characterization of excited carrier relaxation between local conduction band minima in different parts of the BZ.  In Fig.~\ref{fig3}(b), we present the evolution of excited state signals that have been collected with $V_r$ set so as to probe just above the Fermi level. Since every electron with a kinetic energy greater than $eV_r$ is collected by the FeSuMa, all the unoccupied states can be monitored concurrently, regardless of their energy dispersion. A comprehensive discussion of the dynamics, both for bulk and single-layer WSe$_2$, can be found in multiple publications (\textit{e.g.} Refs.~\cite{Bertoni2016, Puppin2022, Madeo2020}). Here, we simply highlight that the FeSuMa allows detection of localised charge populations in a large portion of the BZ simultaneously, allowing for identification of scattering pathways in the material. The time traces in Fig.~\ref{fig3}(c) were collected over 20~mins, corresponding to 36~s of acquisition per frame.  As can be seen in the figure, the statistics are excellent, despite having been acquired with a low probe flux of only 10$^8$ photons/second.

We note certain limitations of the efficient approach just described:  here, time-resolved measurements are performed by integration,  maintaining $V_r$ at a set value. The analysis of data acquired in this way can be challenging if there are multiple excitations at different binding energies but similar $k$; furthermore, access to information about band curvatures is restricted.  Of course, the dataset can be extended to four dimensions ($k_x$, $k_y$, $E$, $\Delta$), simply by sweeping $V_r$ in the manner described above (leading to longer acquisition times).

\begin{figure}
\includegraphics[width=1.0\textwidth]{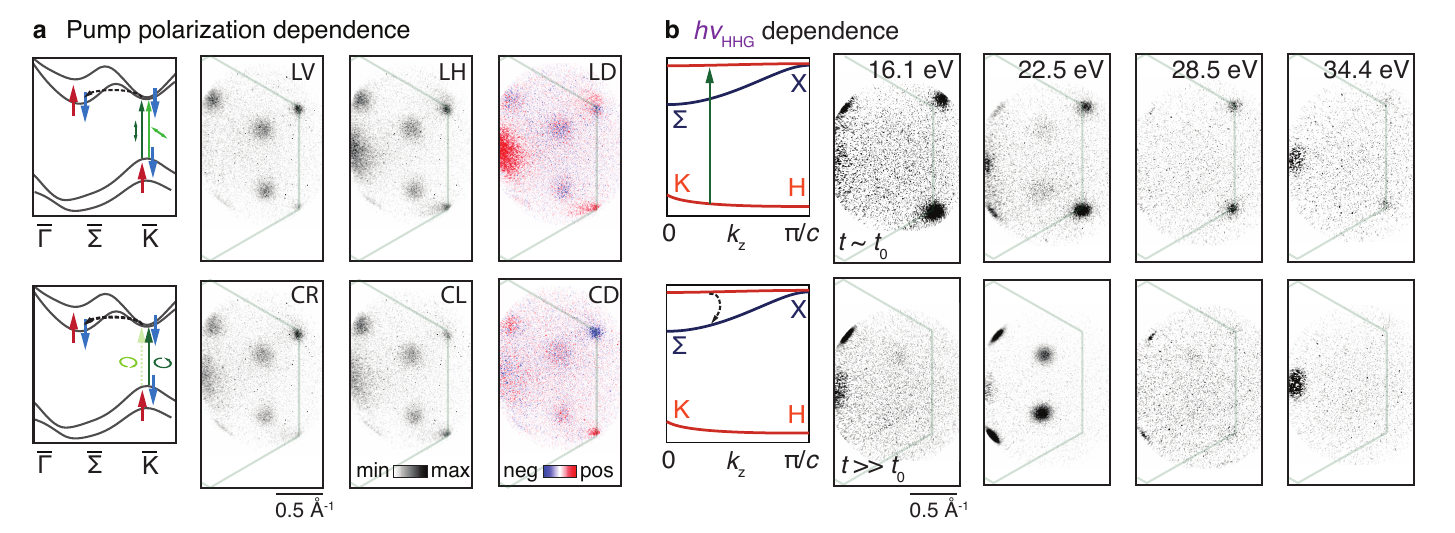}\\
\caption{(a) Spin- and valley-polarised excited carriers in the surface electronic band structure of 2$H$-WSe$_2$ along the {\Gammabar-\Kbar} high-symmetry line. At left, adapted from Ref.~\cite{Bertoni2016}, red and blue color-coding in the band structure plot refers to the spin polarisation of the bands. Light and dark arrows symbolise right- and left-circular polarisation of the pump pulse.  The data shows pump polarization-dependent measurements of excited-state spectra. The labels at the upper right-hand corners indicate linear vertical (LV) and horizontal (LH) polarizations, linear dichroism (LD) as a difference plot of LH-LV, circular right (CR) and left (CL) polarizations, and circular dichroism (CD) as a difference plot of CL-CR. Probe energy was 22.6~eV, and the spectra were collected at the time delay of 200~fs. 
(b) Probe-photon-energy-dependent excited-state spectra at the peak of excitation (top row) and at 200~fs after the excitation (bottom row).  The probe photon energy is indicated at the upper right-hand corner at the top of each column. The out-of-plane electronic dispersion along $\mathrm{\Sigma}$-X leads to a photon-energy-dependence in the photoemission intensity of the states projected into \Sigmabar.  Meanwhile, the states along K-H are nearly non-dispersing, and thus the photoemission intensity in those states is nearly independent of probe photon energy.  The schematic of the out-of-plane-dispersing band structure is adapted from Ref.~\cite{Bertoni2016}.  Data were acquired with $V_r$ set such that $E-E_F = 0.65$~eV.  }
\label{fig4}
\end{figure}

An advantage that the FeSuMa shares with PEEM and momentum microscopy is the capability for maintaining a fixed sample geometry while mapping the momentum space.  Incident light polarisation can remain fixed and photoemission matrix elements unchanged throughout an experiment, and one can straightforwardly extract information such as dichroism from excited-state populations.
In Fig. \ref{fig4}(a), we show the excited carrier distributions that arise in 2$H$-WSe$_2$ pumped with four polarisations:  linear vertical (LV), linear horizontal (LH), circular right (CR), and circular left (CL).  Here, the choices of photon energy and acceptance angle do not image the whole BZ, but allow us to simultaneously see dynamics at the inequivalent \Kbar- and \Kbar'-points and at the corresponding \Sigmabar- and \Sigmabar'-points.  The excitation with linearly polarised light leads to negligible linear dichroic (LD) contrast in the population at \Kbar- and \Sigmabar-points. Pumping with LH light produces a strong signal around the \Gammabar-point; this is a known consequence of multi-photon photoemission process enhanced by this polarisation \cite{Miaja2006, Keunecke2020}.
On the other hand, we see significant circular dichroic (CD) signal at the adjacent \Kbar- and \kbarp-points.  This arises due to a combination of (1) the primarily two-dimensional character of the states at \Kbar- and \kbarp, and (2) the surface sensitivity of the ARPES measurement \cite{Bertoni2016,Riley2014}.  Indeed, the low photoelectron kinetic energies in the measurements described here mean that these spectra are highly sensitive to the physics of the topmost atomic layer of the crystalline structure \cite{Seah1979}.

A full movie of dynamics in a different material system---Bi(111)---is available in the Supplementary Materials \cite{suppMat}.

A powerful aspect of the Artemis setup is its ability to switch efficiently between different HHG probe energies.  (See also Ref.~\cite{Heber2022}.)  This is possible because of carefully optimized optical alignment in the beamline and fine angular control of the final toroidal focusing mirror.  Thus, we can coarsely map the out-of-plane dispersion of unoccupied states, in a manner analogous to that by which the occupied-state $k_z$-dispersion is obtained at synchrotron light sources. We demonstrate this principle in Fig. \ref{fig4}(b). Varying the probe energy leads to strikingly different excited state signals across the BZ. The lowest-lying conduction-band states along the $K$-$H$ path are nearly non-dispersive \cite{Voss1999}, and are visible at all photon energies.  However, the scattering from $K$ to $\Sigma$  is well captured at only one probe energy, 22.4~eV.  In this connection, we note both that the out-of-plane dispersion along the $\Sigma$-$X$ path is more pronounced than that along $K$-$H$ path \cite{Voss1999}, and also that the photoemission matrix elements are presumably enhanced at particular probe energies \cite{Heber2022,Boschini2020,Moser2017}.
The first of these points highlights the importance of thoughtful HHG photon-energy selection in studies of materials in which 3D-dispersing band structures play an important role; this is the case, for example, in Weyl candidates Co$_3$Sn$_2$S$_2$\cite{Liu2021} and PtTe$_2$ \cite{Yan2017}. The second points to the possibility of using matrix element effects to optimise signal-to-noise for all types of samples, including those with a primarily 2D electronic character.

\section{\label{technical}Additional technical considerations}

\begin{figure}
\includegraphics[width=1.0\textwidth]{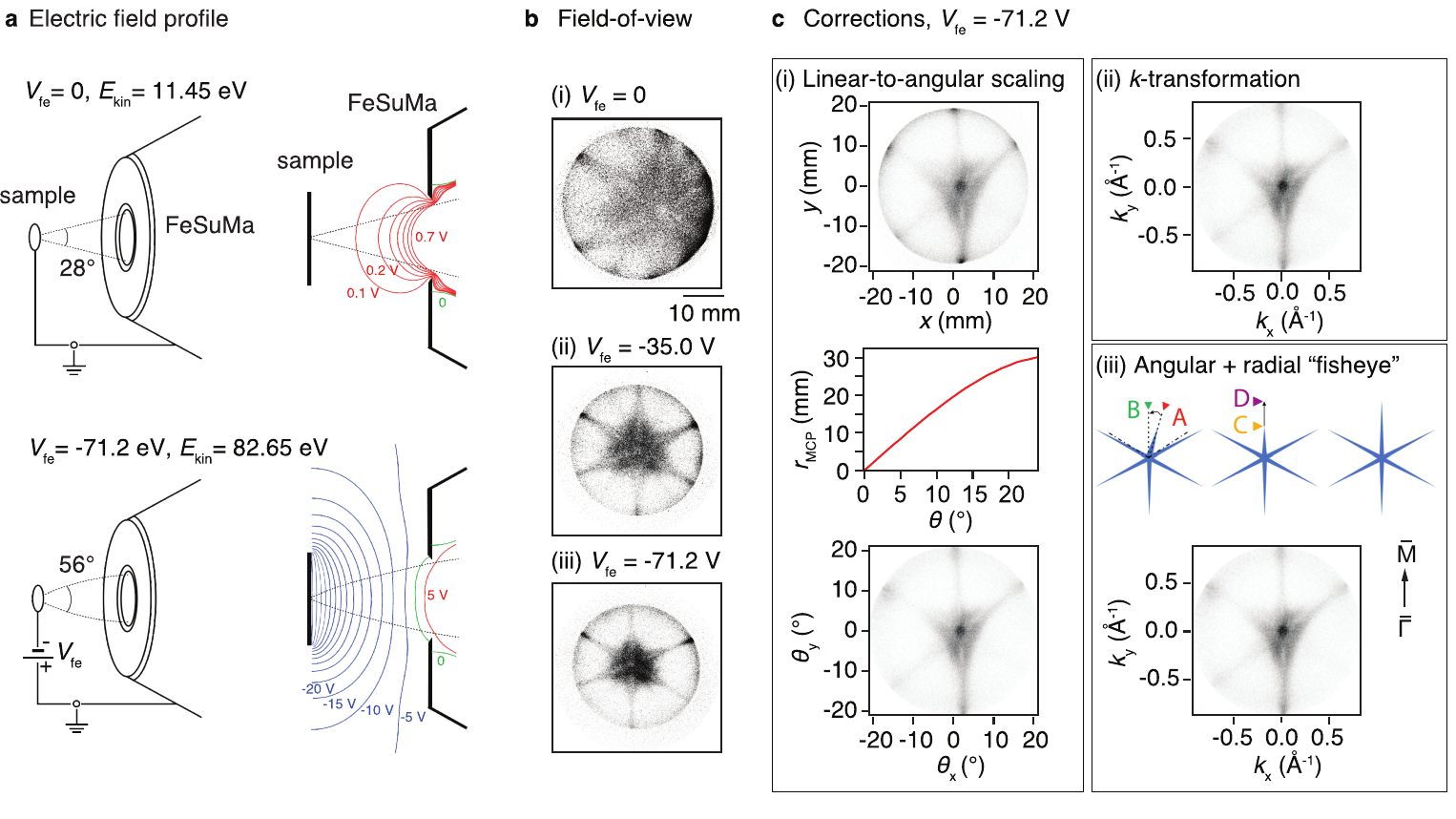}\\
\caption{ (a) Electric fields between sample and FeSuMa analyser for normal operation mode (top) and when "fisheye voltage" is applied (bottom). Calculated with \textsc{SIMION} \cite{Dahl2000}. (b) Fermi surface of Bi(111) taken with probe energy of 16.2~eV, with increasing ``fisheye voltage" applied.  (The focusing conditions were optimized at condition ii, with the result that the focusing conditions are slightly non-optimal in (i) and (iii) and the image appears off-centre on the detector.) (c) Workflow for applying the corrections required for datasets acquired with ``fisheye voltage". (i) Conversion between position on the MCP, $(x, y)$, to emission angle, $\theta_x, \theta_y$, based on the calibration radial function (red curve) obtained from ray tracing. (ii) Angle-to-momentum transformation. (iii) Angular and radial corrections to the image based on the expected symmetry of the intensity distribution. Black dot-dashed lines represent the portions of the image that need correction, while arrows indicate the direction of the corrections.}
\label{fig5}
\end{figure}

\subsection{\label{fisheye}``Fisheye" Data Acquisition}
Applying a bias voltage to the sample holder is a very convenient approach to increase the momentum field of view \cite{Gauthier2021}. Due to the additional component of the field towards the analyser (shown schematically in Fig. \ref{fig5}(a)), electron trajectories are bent, and electrons that initially deviate strongly from the lens axis are nevertheless able to enter the analyser. Thus, using photon energies of only 16.2, 22.2, 28.6, and 34.2 eV, we cover portions of the momentum space at the Fermi level that are much larger than we would otherwise be able to access without the fisheye voltage, achieving radii of 0.81, 0.88, 0.9, and 0.97~\AA$^{-1}$, respectively. The drawback of this approach is that it can lead to distortions resulting from the presence of the electrical field, especially when cylindrical symmetry around the lens axis is broken by the sample's immediate environment (\textit{i.e.,} non-cylindrical sample holder, manipulator shape, cables, etc.). Because we can easily see the momentum distribution ``live'' before acquiring a spectrum, we can take some steps to minimize distortions by adjusting of the geometry of the experiment.  Further processing after the measurement, based on purely symmetry-driven considerations, allows us to eliminate all visible distortions of the angular distribution. This will now be explained.

We introduce two types of corrections to deal with angular and radial distortions, taking as our starting point the known symmetries of our material systems.  In the angular case, we are concerned with a segment of the dataset where there are distortions like those illustrated schematically by black dashed lines in the left panel of Fig. \ref{fig5}(c)(iii). We take the two axes A and B, as indicated by dashed lines leading to the red and green triangles, respectively, in the left panel of Fig. \ref{fig5}(c)(iii). In the affected segment of the data we then shift all points that lie along the A-axis onto the B-axis. For all other points in this segment, a linear interpolation then squeezes the part of the image that lies to the left of B and stretches the part of the image that lies to the right of B.

For the radial correction, we show an illustrative example in Fig.~\ref{fig5}(iii).  In this simple cartoon, we only need to correct one portion of the image that is obviously compressed relative to the others.  Identifying the two points C and D that lie along the same axis (orange and purple triangles in the central panel of Fig. \ref{fig5}(c)(iii)), we perform a linear interpolation such that C is moved onto D, and all other points in a segment are stretched (or squeezed) linearly while keeping the centre of the image intact.

\subsection{\label{spacecharge}Space-charge}
\begin{figure}
\includegraphics[width=1.0\textwidth]{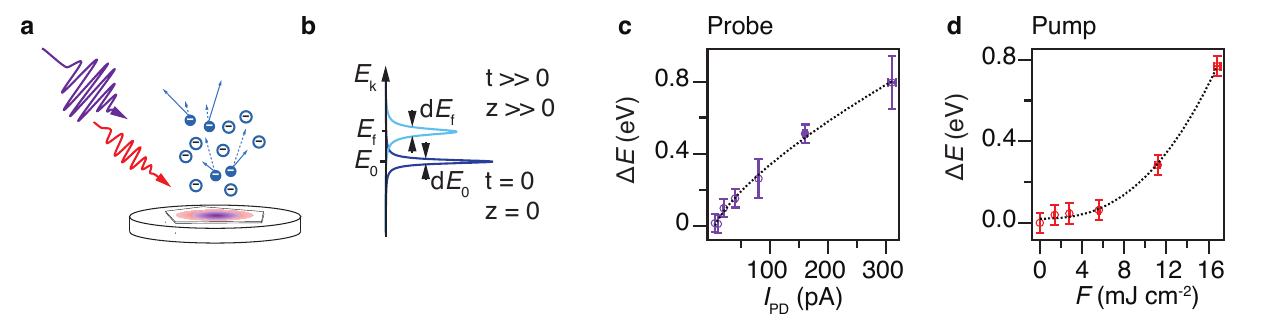}\\
\caption{(a) Cartoon of space-charge generated in a pump-probe photoemission experiment. Coulombic interactions occur between charged particles in a dense cloud of photoemitted electrons. (b) Qualitative impact of space-charge on photoemission spectra. Dark (light) blue peaks represent the electron distribution just after photoemission (after travel towards the analyser). (c) Measured probe-induced peak shift of the spectrum at the peak of excitation, as a function of probe photon flux (measured as photocurrent $I_{PD}$ induced in a photodiode that can be extended into the beampath.  Error bars are estimated from the standard deviation in detected photoelectron counts.) (d) Measured pump-induced peak shift of the Fermi edge before the excitation, as a function of pump fluence. (Horizontal error bars estimated from typical fluctuation in pump power as measured with power meter \cite{suppMat}, vertical error bars determined by uncertainty in fitting the FL shift).}
\label{fig6}
\end{figure}

As sketched schematically in Fig.~\ref{fig6}(a), space charge arises due to Coulombic repulsive interactions within the dense cloud of photoelectrons emitted from the sample surface, leading to energy shifts and distortions of electron trajectories as they move towards the analyser \cite{Hellmann2009,Passlack2006}. The resultant photoemission spectra exhibit reduced effective energy- and momentum-resolution, as well as other artifacts, such as shifting of spectra and possible ``ghost" peaks \cite{Passlack2006}. The energy shift and broadening are illustrated schematically in Fig.~\ref{fig6}(b). In addition to the fact that a dense cloud of Coulombically interacting photoelectrons can be generated by the probe pulse, the pump beam can produce an unwanted cloud of ``slow" secondary electrons via multiphoton photoemission and emission from surface defects \cite{Oloff2016}. 
This latter effect can contribute additional space-charge effects.

In our setup, photoemitted electrons are tightly confined in space and time only once, at the sample surface, before they interact with the MCP \cite{Borisenko2022}. This is an advantageous situation relative to HAs and MMs, where additional focal planes and spatial confinement can cause further Coulombic interaction \cite{,Schonhense2015,Maklar2020,Kutnyakhov2020}.  Moreover, in a ToF, a long-range electric field develops as slow electrons produced by the pump propagate through the lens tube, and fast valence electrons experience an accelerating or decelerating force, depending on the time delay, culminating in a ``fake" time-zero at a large (tens-of-ps) time delay \cite{Kutnyakhov2020}. These effects are largely avoided in the FeSuMa. The retarding voltage readily repels the slowest secondaries---possibly even at the very entrance of the lens column, depending their kinetic energies---so as to reduce their interaction with the other photoelectrons in the lens tube. 

Of course, the severity of distortions always depends also on XUV beam diameter and on pulse energy \cite{Hellmann2012}. At the relatively low 1-kHz repetition rate of the Artemis set-up that was used for this particular experiment, the choice of photon flux was a compromise between the space-charge and the acquisition time required to achieve sufficient signal-to-noise ratio. In future experiments on the Artemis 100-kHz beamline, we expect to see this issue partially remedied.

In Fig. \ref{fig6}(c), we characterise the spectral modifications due to probe-induced space-charge. We use Bi(111) spectra to estimate the shift of the Fermi edge, which occurs across the entire investigated XUV range as a function of flux \cite{Passlack2006}. Fig \ref{fig6}(d) shows the pump-induced Fermi-edge shift. The pump-induced spectral distortions exhibit a complex dependence on the time delay, and are present over a range of several picoseconds after temporal overlap \cite{Ulstrup2015}. The secondary-electron population scales non-linearly with the $n^{th}$-power of the laser fluence and, in general, affects primarily the low kinetic energy portion of a spectrum. These pump-fluence-dependent measurements were made at a time delay of 150~fs before the optical excitation, in order to exclude the affects of real ultrafast dynamics happening in the sample. Up to a fluence of approximately 5~mJ~cm$^{-2}$, the spectra are virtually unaltered by any pump-induced space-charge. Above this threshold, the spectral shift shows a power dependence of $F^x$ ($x = 2.7\pm0.1$), in agreement with a previous study of the excitation with a 1.55~eV pump \cite{Oloff2016}.

\section{\label{conclusions}Conclusions}

The FeSuMa offers a simple, affordable approach to high-quality pump-probe photoemission measurements, particularly for time-resolved ARPES of valence and conduction states near the Fermi level.  Like certain PEEM-based approaches, it permits measurement of dynamics spanning the entire Brillouin zone.  In the context of an HHG beamline that permits scanning of the probe energy, the fisheye mode of operation offers particular benefit for studies of 3D-dispersing states.  The FeSuMa is highly complementary to hemispherical analyzers, and constitutes an attractive option for laboratory-scale measurements of electron dynamics.  Measurements of conduction-band dynamics in layered 2H-WSe$_2$ yield excellent agreement with previously published results based on momentum microscopy.

\begin{acknowledgments}
We thank Phil Rice, Alistair Cox, and the CLF Engineering Section for technical support; and Drs. James O. F. Thompson and Marco Bianchi for helpful discussion.  We acknowledge funding from VILLUM FONDEN through the Centre of Excellence for Dirac Materials (Grant No. 11744) and from the Independent Research Fund Denmark  (Grant No. 1026-00089B).  Work at the Artemis Facility is funded by the UK Science and Technology Facilities Council.  The research leading to these results has received funding from LASERLAB-EUROPE (grant agreement no. 871124, European Union’s Horizon 2020 research and innovation programme).

Supplementary material is available online.  It includes a movie of data acquired with the FeSuMa across a range of delay times before and after an optically pumped excitation in Bi(111)/Bi$_2$Se$_3$, and
information about the time resolution and laser stability on the beamline used for this experiment.

\end{acknowledgments}


\bibliography{FeSuMa}

\end{document}



\title{Supplementary Materials to ``Accessing the full 3D Brillouin zone plus time resolution, with high harmonic generation and a new approach to detection''} 

\author{Paulina Majchrzak}
\affiliation{Department of Physics and Astronomy, Interdisciplinary Nanoscience Center, Aarhus University, 8000 Aarhus C, Denmark}%

\author{Yu Zhang}
\affiliation{Central Laser Facility, STFC Rutherford Appleton Laboratory, Research Complex at Harwell, Harwell, OX11 0QX, United Kingdom}

\author{Andrii Kuibarov}
\affiliation{Leibniz IFW Dresden, Helmholtzstr. 20, 01069 Dresden, Germany}

\author{Richard Chapman}
\affiliation{Central Laser Facility, STFC Rutherford Appleton Laboratory, Research Complex at Harwell, Harwell, OX11 0QX, United Kingdom}

\author{Adam Wyatt}
\affiliation{Central Laser Facility, STFC Rutherford Appleton Laboratory, Research Complex at Harwell, Harwell, OX11 0QX, United Kingdom}

\author{Emma Springate}
\affiliation{Central Laser Facility, STFC Rutherford Appleton Laboratory, Research Complex at Harwell, Harwell, OX11 0QX, United Kingdom}

\author{Sergey Borisenko}
\affiliation{Leibniz IFW Dresden, Helmholtzstr. 20, 01069 Dresden, Germany}

\author{Bernd B\"uchner}
\affiliation{Leibniz IFW Dresden, Helmholtzstr. 20, 01069 Dresden, Germany}

\author{Philip Hofmann}
\affiliation{Department of Physics and Astronomy, Interdisciplinary Nanoscience Center, Aarhus University, 8000 Aarhus C, Denmark}%

\author{Charlotte E. Sanders}
\affiliation{Central Laser Facility, STFC Rutherford Appleton Laboratory, Research Complex at Harwell, Harwell, OX11 0QX, United Kingdom}%
 \email{charlotte.sanders@stfc.ac.uk.}

\date{\today}



\maketitle 


\section{\label{sec:level2}Sample Preparation}

The Bi(111) surface was prepared by Bi deposition from a homemade evaporator onto a freshly cleaved Bi$_2$Se$_3$ crystal in vacuum conditions of low-$10^{-9}$~mbar.  We estimated the growth rate as approximately 0.7--0.8~Å/s, on the basis of calibration with a quartz crystal monitor (Inficon XTC/3 with UHV bakeable sensor).  After approximately 20~nm of deposition, we gently annealed the film at less than circa 250\textdegree C for 10~min.  Then we repeated the growth procedure.  We estimated the final Bi film thickness at approximately 37~nm.

The growth chamber is connected directly to the analysis chamber (separated by a gate valve), for fully \textit{in-vacuo} transfer.

The 2$H$-WSe$_2$ bulk crystal was obtained commercially from HQ Graphene. The sample was cleaved \textit{in-situ} with a Kapton tape, at a pressure of $2\times10^{-9}$~mbar.

\newpage
\section{\label{sec:level1}Beamline Time Resolution}

In Fig.~\ref{sfig1}, we show an auto-correlation trace of the fundamental beam obtained by TiPA (Light Conversion Ltd.). The full-width half-maximum (FWHM) of the auto-correlation peak is approximately 57~fs, which corresponds to a Gaussian pulse width of about 40~fs.\\

\begin{figure}[!h]
\includegraphics[width=0.70\textwidth]{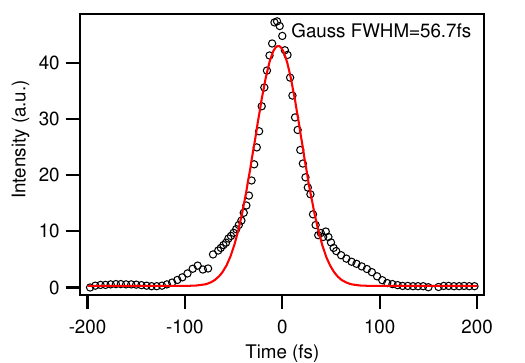}\\
\caption{The auto-correlation measurement of the fundamental beam (from a TiPA, Light Conversion Ltd). The FWHM of the Gaussian fit gives 57 fs.}
\label{sfig1}
\end{figure}

\newpage
\section{\label{sec:level2}Pump Stability}
The pulse-to-pulse stability of this laser system is typically around 2\%. The stability of the power, averaged over a second---which is more relevant than pulse-to-pulse stability to time-integrated pump-probe measurements---is generally around 0.4\%. We note, however, that the stability of the pump is not critical to the results of the present study.\\

\begin{figure}[!h]
\includegraphics[width=0.70\textwidth]{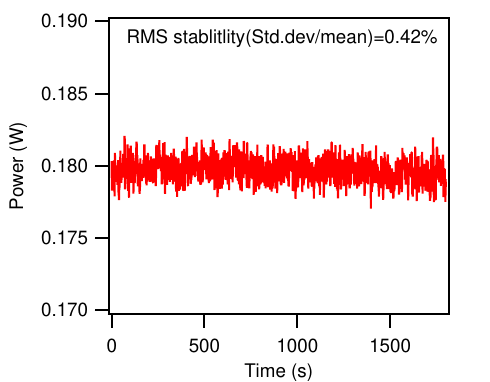}\\
\caption{Representative measurement of laser stability, as  measured on the pump beam by a power meter (UP19S-15S-H5, Gentec-EO, Inc.) over a period of 30 min. Each data point is the average of 6 readings over the course of 1~s. The root mean square (RMS) stability is 0.4\%. }
\label{sfig1}
\end{figure}
